\documentclass[a4paper]{jpconf}
\usepackage{graphicx}
\usepackage{makecell}
\usepackage{booktabs}
\usepackage{hyperref}

\begin{document}

\begin{center}
{Preprint accepted at CISBAT 2023 - The Built Environment in Transition, Hybrid International Conference, EPFL, Lausanne, Switzerland, 13-15 September 2023}
\end{center}

\title{The Building Data Genome Directory -- An open, comprehensive data sharing platform for building performance research}

\author{Xiaoyu Jin$^{1}$, Chun Fu$^{2}$, Hussain Kazmi$^{3}$, Atilla Balint$^{3}$, Ada Canaydin$^{3}$, Matias Quintana$^{2, 4}$, Filip Biljecki$^{2}$, Fu Xiao$^{1}$, Clayton Miller$^{2,*}$}
\address{$^1$ Department of Building Environment and Energy Engineering, The Hong Kong Polytechnic University, Hong Kong}
\address{$^2$ College of Design and Engineering, National University of Singapore (NUS), Singapore}
\address{$^3$ Department of Electrical Engineering, KU Leuven, Belgium}
\address{$^4$ Future Cities Laboratory Global, Singapore-ETH Centre, Singapore}
\ead{$^*$clayton@nus.edu.sg}


\begin{abstract} 
The building sector plays a crucial role in the worldwide decarbonization effort, accounting for significant portions of energy consumption and environmental effects. 
However, the scarcity of open data sources is a continuous challenge for built environment researchers and practitioners. 
Although several efforts have been made to consolidate existing open datasets, no database currently offers a comprehensive collection of building data types with all subcategories and time granularities (e.g., year, month, and sub-hour).
This paper presents the Building Data Genome Directory, an open data-sharing platform serving as a one-stop shop for the data necessary for vital categories of building energy research. 
The data directory is an online portal (\href{http://buildingdatadirectory.org/}{\url{buildingdatadirectory.org/}}) that allows filtering and discovering valuable datasets.
The directory covers meter, building-level, and aggregated community-level data at the spatial scale and year-to-minute level at the temporal scale. 
The datasets were consolidated from a comprehensive exploration of sources, including governments, research institutes, and online energy dashboards.
The results of this effort include the aggregation of 60 datasets pertaining to building energy ontologies, building energy models, building energy and water data, electric vehicle data, weather data, building information data, text-mining-based research data, image data of buildings, fault detection diagnosis data and occupant data. 
A crowdsourcing mechanism in the platform allows users to submit datasets they suggest for inclusion by filling out an online form. 
This directory can fuel research and applications on building energy efficiency, which is an essential step toward addressing the world's energy and environmental challenges. 
\end{abstract}

\section{Introduction}
The rise of artificial intelligence as a tool for built environment applications has the potential to impact several industries significantly.
However, data availability in the built environment domain remains a critical bottleneck due to privacy concerns and acquisition costs \cite{Jin2022review}.
Open data sources are essential for understanding energy consumption patterns, identifying areas for improvement, and testing energy-saving strategies, especially in the absence of in situ measurements. 
Yet, access to open data sources in the built environment domain lags behind other communities \cite{kazmi2021towards}, posing limitations for researchers and practitioners in developing effective energy-saving solutions \cite{FJohari2020Urban}.
In addition to limited accessibility, available open datasets are often dispersed and require labor-intensive and time-consuming collation due to varying formats and sources \cite{JIN2022120210}. 
Efforts have been made to aggregate open datasets and share them through platforms or directories such as the Building Performance Database (BPD) \cite{Walter2016-xy}, the Building Data Genome (BDG) projects \cite{miller2017building, miller2020building}, and the Directory of Buildings Energy Consumption Datasets (DBECD) \cite{babaei2015study}. 
However, these projects have limitations in the diversity of data types, lack of user contributions, and missing data.

This paper outlines the development of a comprehensive data-sharing platform for building performance research. 
This effort is achieved by creating a data directory that is publicly available and includes functions for filtering, visualization, and uploading new data sets. 
The Building Data Genome Directory is a lightweight web app that links to a wide range of open datasets, offering users easy access to comprehensive coverage of relevant information. 
In subsequent sections, the paper will introduce the data sources, data category definitions, reasons for inclusion, critical functions of the web app, and some application cases.

\section{Data sources}
\begin{figure}
    \centering
    \includegraphics[width=1.05\linewidth]{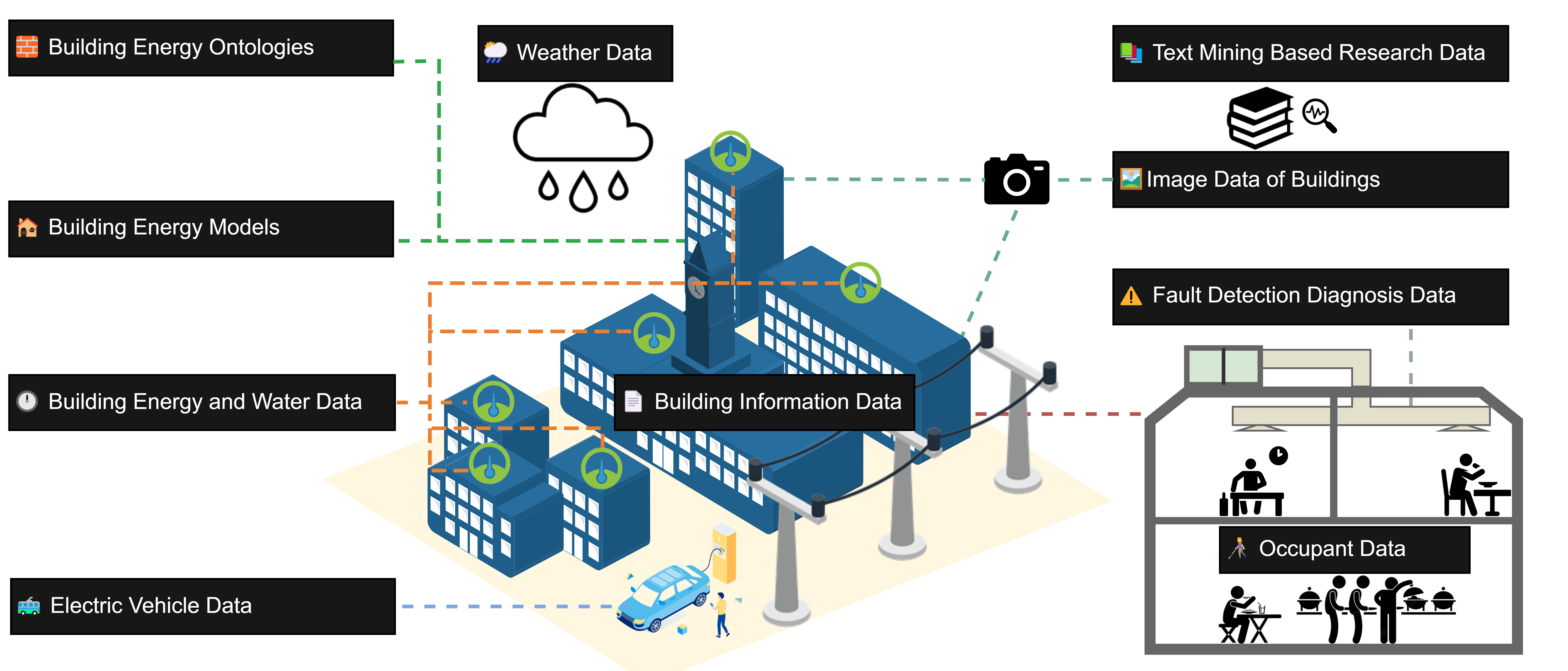}
    \caption{Schematic of the categories of datasets included in Building Data Genome Directory}
    \label{fig:types}
\end{figure}

The directory focuses on collecting information about open building performance datasets that are widely dispersed and fragmented, which conventionally would require a rigorous data collection process. 
Metadata for the directory was gathered from various open data sources, including government disclosure programs, research projects, institutes, and publicly available dashboards. 
Details on each of these data source categories are discussed in the following subsections.
The directory data sources are divided according to category and type of data based on the format (e.g., tabular, image) and process of the system that created the data (e.g., HVAC, occupants, sensors).
Figure \ref{fig:types} shows an overview of the data set categories, which will be outlined in the following subsections.

\subsection{Government disclosure data}
Data from government disclosure programs is a significant source for built environment data. 
One example is the Local Law 84 (LL84) of New York City (NYC) in the United States, which requires building owners to disclose their energy and water consumption data through benchmarking annually \cite{kontokosta2012predicting}.
This directive has led to the publication of the Energy and Water Data Disclosure dataset for Local Law 84 by the NYC government.
These city-level datasets can contain many samples, with some featuring tens of thousands of buildings, although they may have coarse-grained time intervals of a year or a month.
To collect these datasets, a comprehensive review of relevant literature and examination of laws pertaining to data disclosure was conducted \cite{Jin2022review}. 
Open data portals provided by city governments~\cite{2023_ijgis_3d_city_index}, such as the NYC open data portal (\url{https://opendata.cityofnewyork.us/}), were also browsed to gather available datasets, ensuring the comprehensiveness of the data directory.

\subsection{Open research data}
Research institutes and organizations have published various datasets for building performance research. 
Some datasets are available on websites, such as the Building Data Genome dataset on Kaggle \cite{miller2017building} or the 3D city model of Singapore public housing buildings on GitHub \cite{2020_3dgeoinfo_3d_asean}. 
Other datasets are published through journals, with \textit{Scientific Data} being a significant venue. 
A recent review has also listed open-source datasets for building energy demand \cite{kazmi2021towards}.
These datasets typically provide detailed information about individual buildings but may not have large numbers of samples (generally less than 5,000). 
A common differentiator of these types of data sets is that the time-series frequency may be higher, sometimes even at the minute level, offering a more granular view of a building's energy usage. 
Some datasets also provide detailed information about building characteristics, solar installations~\cite{2021_land_roofpedia}, morphological indicators~\cite{2022_ceus_gbmi}, or sensor locations and building structure \cite{luo2022three}. 
Accessing and leveraging these datasets allows researchers to gain comprehensive insights into individual buildings and their energy usage.
To collect these datasets, relevant reviews and research papers were examined, including platforms that provide access to datasets referenced in articles, 

\subsection{Data collected from open, online dashboards}
In response to the growing emphasis on net-zero and sustainability goals in the higher education sector, many educational institutes and universities, such as the University of California, Berkeley, Cornell University, and Princeton University, have public energy management dashboards that provide access to energy usage data for further study and analysis.
For these datasets, a data acquisition pipeline can be built using scripts to automate the process of extraction from these dashboards, enabling batch downloads of performance data from thousands of buildings. 
The directory includes several datasets that were retrieved from these types of public web-based energy management dashboards.
For many of these dashboards, the API of the data source can usually be found using built-in web browser developer tools. 
Once the data API is identified, an automated process can be configured with the required data parameters, such as building ID and specific time period, to enable batch downloading of performance data from a web-based dashboard.

\begin{table}[]
\begin{footnotesize}

\caption{Categories of the data in the directory with short descriptions and an example representative dataset of each type.}
\begin{tabular}{p{0.23\textwidth}p{0.39\textwidth}p{0.3\textwidth}}
\toprule
\textbf{Category   Name}          & \textbf{Data Scope Description}                                                                   & \textbf{Representative Dataset}                                                                                 \\ \midrule 
Building   Energy Ontologies      & Haystack, Brick Schema, and   datasets supporting the semantic model development                  & A Synthetic Building Operation   Dataset \cite{li2021synthetic}                                \\
Building   Energy Models          & Building energy model   information and simulation data of prototype buildings                    & Building Stock dataset   \cite{Frick2019-kh}                                                           \\
Building   Energy and Water Data  & Time-series energy data and   water consumption data of buildings                                 & Energy and Water Data Disclosure   for Local Law 84 \cite{kontokosta2012predicting}                                \\
Electric   Vehicle Data           & Charging infrastructure datasets   and time-series energy consumption data from charging sessions & Campus   Electric Vehicle Charging Stations Behavior                                                            \\
Weather   Data                    & Historical data from environmental sensors and weather prediction                               & Campus Electric Vehicle Charging   Stations Behavior \cite{EV}                                 \\
Building   Information Data       & Stock characteristics, GIS data,   and project management data for a large number of buildings     & European Building Stock   Characteristics Dataset \cite{milojevic2023eubucco}               \\
Image   Data of Buildings         & Image data of a large number of   buildings                                                       & Annotated   Image Database of Architecture \cite{2021_caadria_aida}                         \\
Text   Mining Based Research Data & Text-mining data from previous   research and communities                                         & A Comprehensive Text-mining   Driven Review of Scientific Literature \cite{abdelrahman2021data} \\
Fault Detection and Diagnosis Data  & Ground-truth and simulated datasets for anomalies in the built environment and building systems                               & Large-scale Energy Anomaly   Detection (LEAD) Dataset \cite{Fu2022-nq}                              \\
Occupant   Data                   & The thermal comfort data of   occupants collected from experiments                                & Cozie smartwatch application \cite{cozie}                                                    \\ \bottomrule
\end{tabular}
\label{tab:sources}
\end{footnotesize}

\end{table}


\section{Overview of the directory interface} 
The Building Data Genome Directory can be found online at: \href{http://buildingdatadirectory.org/}{\url{buildingdatadirectory.org/}}.
The interface comprises of a main page, referred to as the \emph{Meta Directory}, which provides an overview of all available datasets and several sub-pages presenting datasets by types. 
The \emph{Meta Directory} page introduces the Building Data Genome Directory and outlines the scope of the collected datasets. 
As a web app, it has filtering, visualization, and uploading functions for the datasets. 
Datasets pertaining to buildings, such as \emph{Building Energy and Water} and \emph{Building Information} provide geospatial granularity levels that correspond to individual buildings or, at the very least, communities, instead of the aggregated data of an entire city.
The \emph{Meta Directory} includes a schematic diagram showcasing the various datasets available in the Building Data Genome Directory, as shown in Figure \ref{fig:types}. 
Each black label in the diagram represents a specific data type and has a corresponding subpage, with its link conveniently located on the left column of the web page. 
The scope description for these types and the representative datasets are presented in Table \ref{tab:sources}.
The \emph{Add New Dataset} uploading function is at the bottom of the left-hand button. 
Users must fill in the \emph{Dataset Name}, \emph{URL}, and \emph{Dataset Type} items to submit a possible contribution to the directory. 
The datasets submitted by the users will be stored and displayed at the bottom of the \emph{Meta Directory} page, and they will be added to the directory after undergoing a review process.

\begin{figure}
    \centering
    \includegraphics[width=1.05\linewidth]{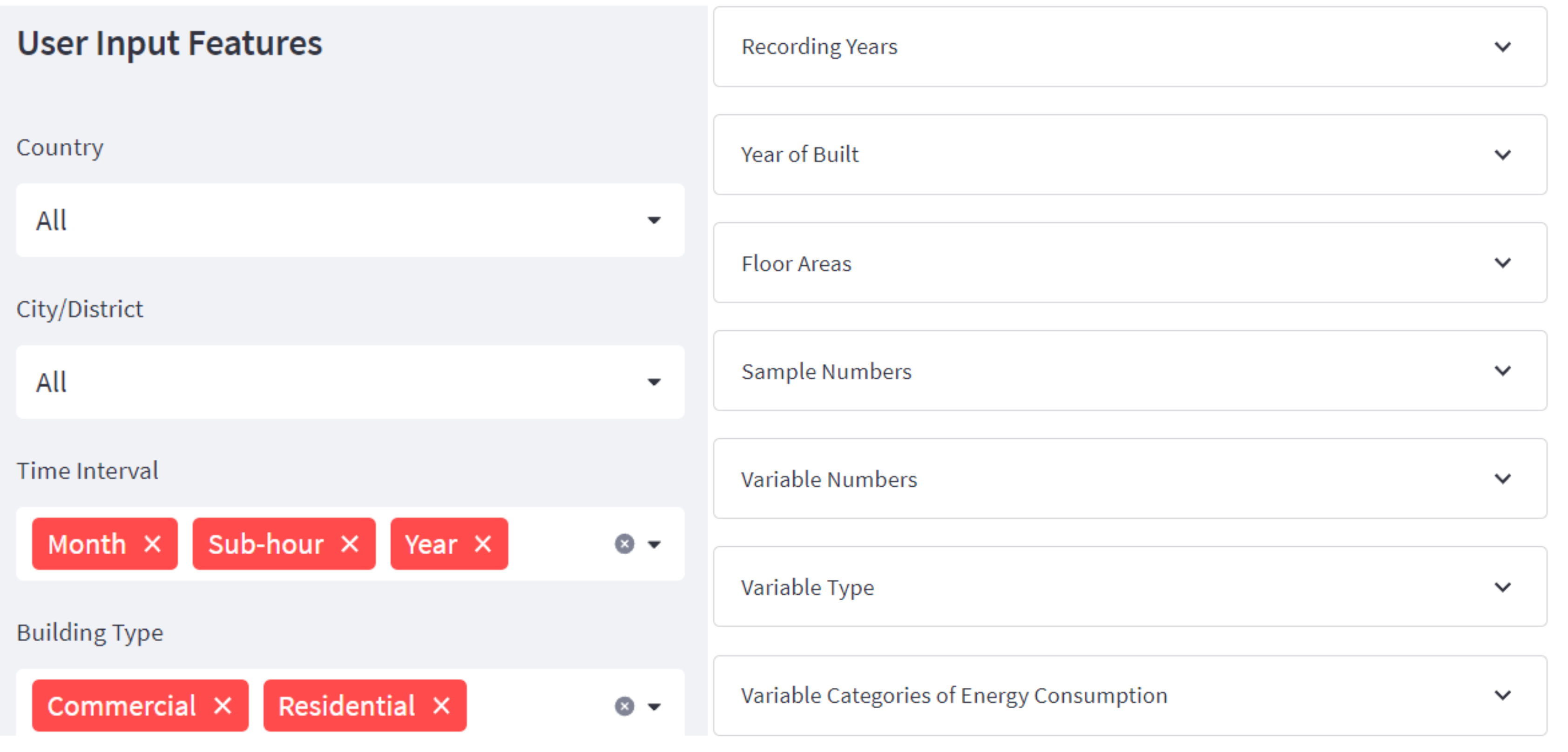}
    \caption{Building Data Genome Directory interface showcasing the filtering and visualization functions.}
    \label{fig:functions}
\end{figure}

The category with the highest number of data sets is \emph{Building Energy and Water}, which includes over 30 datasets at the moment. 
A metadata table that provides essential information about the datasets is displayed on this page, including disclosure status (e.g., data opening level, license availability, organization) and information on the building samples. 
Figure \ref{fig:functions} shows the filtering and visualization functions.
The filtering functions enable users to select datasets by location, time interval, and building type. 
The visualization functions include bar plots with adjustable axes to visualize numerical information, bubble plots to display sample and variable numbers with the size of circles denoting sample sizes and variable quantities, and heatmaps to visualize variable categories.

\section{Conclusion and future works}
The Building Data Genome Directory is a potentially valuable resource for building energy research, providing comprehensive datasets and web app functions for filtering, visualization, and uploading. 
This directory can be a starting point for researchers and analysts who want to start the exploration process for applicable open data sets for their studies.
Numerous research endeavors are anticipated to emerge as branches stemming from this directory. 
As highlighted by Jin et al.~\cite{Jin2022review}, the availability of comprehensive datasets will significantly expedite research in building energy, encompassing areas such as building energy management, grid management, and socio-economic analysis. 
The team is developing a sub-branch within the Building Data Genome Directory focusing on time-series feature analysis utilizing energy consumption data.

\subsection{Future expansion and data quality considerations}
Future work can optimize the directory by improving functions such as allowing brief dataset descriptions during uploading and incorporating semantic searching capabilities. 
Enhancing search capabilities for different data types, such as geographic location, would also improve usability, as well as considering unconventional data sources such as scraping relevant data on buildings from property websites~\cite{2022_ui_real_estate_mining} and considering volunteered geographic information such as OpenStreetMap~\cite{2023_bae_osm_qa} in locations that have data of reliable quality.
Finally, to strengthen the crowdsourcing aspect of our platform, we plan to implement a functionality to allow users to flag erroneous information and allow trusted users to edit the database.
Building a community around the directory would foster user communication and optimize the web app. 
Collecting feedback and insights through discussions and forums would provide valuable inputs for enhancing features and usability. 
By actively engaging with users, the directory can continue to evolve and serve as a valuable resource for building energy researchers
b

\section*{Acknowledgments}
The authors gratefully acknowledge the support for this research from the National Key Research and Development Program of China (2021YFE0107400), the Research Grants Council of the Hong Kong SAR (C5018-20GF), and the Singapore Ministry of Education (MOE) Tier 1 Grants: A-0008301-01-00 and A-8000139-01-00.

\section*{References}
\bibliographystyle{elsarticle-num} 
\bibliography{references} 

\end{document}